\documentclass[aps,prl,superscriptaddress,twocolumn]{revtex4-1}

\newcommand{\marker}[1]{#1}
\usepackage{graphicx}
\newcommand{\B}{\mathbf}
\usepackage{amsmath}
\usepackage{amssymb}
\usepackage{verbatim}
\newcommand{\tens}{\B}
\renewcommand{\d}{\!\mathrm{d}}

\newcommand{\frob}{\odot}
\usepackage{bm}
\usepackage{booktabs}
\begin{document}
\title{Inverse design of light-matter interactions}

\author{Robert Bennett}
\affiliation{Physikalisches Institut, Albert-Ludwigs-Universit\"at
Freiburg, Hermann-Herder-Str. 3, 79104 Freiburg, Germany}
\affiliation{School of Physics \& Astronomy, University of Glasgow, Glasgow, G12 8QQ, United Kingdom}

\author{Stefan Yoshi Buhmann}
\affiliation{Physikalisches Institut, Albert-Ludwigs-Universit\"at
Freiburg, Hermann-Herder-Str. 3, 79104 Freiburg, Germany}
\date{\today}

\begin{abstract}
Inverse design represents a paradigm shift in the development of nanophotonic devices, where optimal geometries and materials are discovered by an algorithm rather than symmetry considerations or intuition. Here we present a very general formulation of inverse design that is applicable to atomic interactions in external environments, and derive from this some explicit formulae for optimisation of spontaneous decay rates, Casimir-Polder forces and resonant energy transfer. Using the Purcell factor of latter as an example, we use finite-difference time-domain techniques to demonstrate the ability of inverse design algorithms to go far beyond what can be achieved by intuition-based approaches, opening up a new route to their technological exploitation. 
\end{abstract}

\newcommand{\spontDecayF}{$\begin{aligned}& (2\mu_0\omega^2/\hbar) \, \B{d}_\text{A} \cdot \text{Im} \tens{G}(\B{r},\B{r}',\omega) \cdot \B{d}_\text{A} \\ & \qquad \times \delta(\B{r}-\B{r}_\text{A}) \delta(\B{r}'-\B{r}_\text{A})  \delta(\omega - \omega_\text{A})\end{aligned}$}

\newcommand{\spontDecayDF}{$\displaystyle{\frac{2\mu_0^2\alpha n\omega_\text{A}^4}{\hbar} \, \text{Im}\Big\{ [   \B{d}_\text{A}\cdot \tens{G}^\mathrm{T}(\B{s},\B{r}_\text{A},\omega_\text{A}) ]\cdot [  \tens{G}(\B{s},\B{r}_\text{A},\omega_\text{A} )\cdot \B{d}_\text{A}]}\Big\}$}

\newcommand{\RETF}{$\begin{aligned}&({2\pi \mu_0^2 \omega^4}/{\hbar}) \left| \B{d}_\text{A} \cdot \tens{G}(\B{r}, \B{r}',\omega)  \cdot \B{d}_\text{D} \right|^2 \notag \\
& \qquad\times  \delta(\B{r}-\B{r}_\text{A}) \delta(\B{r}'-\B{r}_\text{D})  \delta(\omega - \omega_\text{D})\end{aligned}$ }

\newcommand{\RETDF}{$\begin{aligned}\frac{4\pi\alpha n\mu_0^3 \omega_\text{D}^6}{\hbar}  \text{Re} \Big\{   \B{d}_\text{A}  \cdot  & \tens{G}^*(\B{r}_\text{A}, \B{r}_\text{D},\omega)\cdot  \B{d}_\text{D} \notag \\
 &\times  [ \B{d}_\text{A}\cdot \tens{G}^\text{T}(\B{s},\B{r}_\text{A}) ]\cdot [ \tens{G}(\B{s},\B{r}_\text{D})\cdot \B{d}_\text{D}] \Big\}\end{aligned}$}

\newcommand{\CasPolF}{$\begin{aligned}\displaystyle{\frac{\mu_0}{\pi} \int_0^\infty \frac{ \d\omega \, \omega^2}{\omega_\text{A}+\omega}   \B{d}_\text{A} \cdot [\nabla  \tens{G}(\B{r},\B{r}',\omega)] \cdot \B{d}_\text{A}}\\
\qquad\times  \delta(\B{r}-\B{r}_\text{A}) \delta(\B{r}'-\B{r}_\text{A}) \end{aligned} $}

\newcommand{\CasPolDF}{$\begin{aligned} \frac{\mu_0^2\alpha n}{\pi} \text{Im} \int_0^\infty & \frac{ \d\omega \, \omega^2}{\omega_\text{A}+\omega}  [\B{d}_\text{A}\cdot \tens{G}^\mathrm{T}(\B{s},\B{r}_\text{A},\omega) \overleftarrow{\nabla} ]\cdot [  \tens{G}(\B{s},\B{r}_\text{A},\omega)\cdot \B{d}_\text{A}] \end{aligned} $}

\maketitle

Traditional design methods work by specifying a device, then investigating its properties. By contrast, in inverse design the desired property is specified, and an algorithm is left to find a device which fulfils the desired criteria. A naive approach to this would be simply trying all devices that fulfil some set of design constraints. The large space of possible designs renders this numerically unrealistic, meaning that a pre-determined set of designs must be optimised over, at least in the earliest applications of inverse methods to electromagnetic problems \cite{Spuhler1998, Dobson1999}. The development of adjoint methods  \cite{Jameson1988} originally used in aerodynamics have made unconstrained inverse design computationally feasible, with the first application in photonics being to low-loss waveguide bends \cite{Jensen2004}. Adjoint methods were subsequently applied to band gaps \cite{Kao2005}, solar cells \cite{Alaeian2012}, on-chip demultiplexers \cite{Piggott2015} and many more diverse systems --- see the recent review articles \cite{Jensen2011,Molesky2018} and references therein. 

An area in which inverse design has not yet been applied is virtual-photon mediated interactions, such as Casimir-Polder \cite{Casimir1948a} forces and resonant energy transfer \cite{Forster1948}. These processes can be described within a very general formalism known as macroscopic quantum electrodynamics (QED) \cite{Gruner1996a}, where they can be reduced to various functionals of the classical dyadic Green's tensor $\tens{G}$ for a source at $\B{r}'$, observation point at $\B{r}$ and frequency $\omega$ defined to satisfy
\begin{equation}\label{GDefinition}
\nabla \times \nabla \times \tens{G}(\B{r},\B{r}',\omega) - {\frac{\omega^2}{c^2} \varepsilon(\B{r},\omega)}\tens{G}(\B{r},\B{r}',\omega) = \mathbb{I}\delta(\B{r}-\B{r}')\,. 
\end{equation}
subject to given boundary conditions. \marker{The quantised electromagnetic fields can be found from this, for example the electric field in a region with permittivity $\varepsilon(\B{r},\omega)$ and unit permeability is given by $\hat{\B{E}}(\B{r}) = i\int_0^\infty d\omega \int d^3 \B{r}' \frac{\omega^2}{c^2} \sqrt{\frac{\hbar}{\pi \varepsilon_0}\text{Im}\varepsilon(\B{r}',\omega)}\tens{G}(\B{r},\B{r}',\omega) \cdot \hat{\B{b}}(\B{r}',\omega)+\text{h.c.}$ \cite{Gruner1996a}, where $\B{b}^\dagger, \B{b}$ are a set of bosonic creation and annihilation operators for the medium-assisted quantised electromagnetic field}\footnote{These are usually called $\B{f}$ and $\B{f}^\dagger$ in macroscopic QED, here we avoid that notation in order to avoid confusion with the merit functions $f$ introduced later.}. The Green's tensor $\tens{G}$ takes  into account both the geometry and material response of an arbitrarily-shaped medium, meaning that an optimal geometry for particular $\B{r},\B{r}'$ and $\omega$ is represented by a particular functional form of $\B{G}$. It follows that $\B{G}$ is the fundamental object which is to be worked with in inverse design of macroscopic QED. 

In this Letter we begin by introducing the underlying formulae for inverse design of light-matter interactions. We then use the specific example of resonant energy transfer combined with finite-difference time domain (FDTD) techniques to demonstrate that the efficiencies achievable in this method are far beyond those found from `by-hand' constructions, opening up a new direction in the design of \emph{any} light-matter interaction dependent device.   

\paragraph*{General formulation.} In order to carry out any optimisation, we need to define a merit function $F$ which we intend to maximise. In traditional presentations of adjoint optimisation, this function is taken to depend on the electromagnetic fields $\B{E},\B{D},\B{B}$ and $\B{H}$, but all of these are of course deducible from the dyadic Green's tensor so we consider $F$ as being dependent on only $\tens{G}(\B{r},\B{r}',\omega)$. The merit function should be an observable quantity, so we take it to be a real-valued functional of $\tens{G}(\B{r},\B{r}',\omega)$, integrated over all its arguments:
\begin{equation}
F = \int \d^3\B{r} \! \int \d^3 \B{r}' \! \int_0^\infty \!\!\d\omega f[\B{G}(\B{r},\B{r}',\omega)].
\end{equation}
The integrals allow us to take into account a delocalised source and extended observation volume, as well as multi-mode effects. The entries of the tensor $\tens{G}$ are  in general complex-valued, so in principle one could consider variations in the real and imaginary parts separately. However, it is more convenient to consider the complex tensors $\tens{G}$ and $\tens{G}^*$ as independent, in which case the variation of the merit function with $\tens{G}$ is;
\begin{equation}
\delta F = 2\! \int \d^3\B{r} \! \int \d^3 \B{r}'  \int_0^\infty \!\!\!\!\d\omega\text{Re} \left[ \frac{\partial f}{\partial \tens{G}}(\B{r},\B{r}',\omega) \frob \delta \tens{G}(\B{r},\B{r}',\omega)\right]
\end{equation}
where $\frob$ represents the Frobenius product ($\B{A}\frob \B{B} = A_{ij}B_{ij}$) and $\delta \tens{G}$ is a change in the Green's function brought about by an infinitesimal change in the environment. If this change can be considered as being confined to a small volume $V$ containing a number density $n(\B{r}'')$ of atoms with polarisabilities $\alpha(\B{r}'')$, we can write $\B{G}$ in terms of the following Born series;
\begin{align}
\delta \tens{G}(\B{r},\B{r}',\omega )= \mu_0 \omega^2& \!\!\int_V \d^3  \B{r}''  n(\B{r}'')\alpha(\B{r}'') \notag \\
&  \times \tens{G}(\B{r},\B{r}'',\omega ) \cdot \tens{G}(\B{r}'',\B{r}',\omega),
\end{align}
 where $\mu_0$ is the vacuum permeability. The change in the merit function is then given by;
\begin{align} \label{BasicDF}
\delta F &= 2 \mu_0\text{Re} \int \d\omega \, \omega^2 \!\!\int \d^3\B{r} \! \int \d^3 \B{r}'  \! \int_V \d^3 \B{r}'' n(\B{r}'')\alpha(\B{r}'') \notag \\
& \qquad \times \frac{\partial f}{\partial \tens{G}}(\B{r},\B{r}',\omega)\frob \tens{G}^\mathrm{T}(\B{r}'',\B{r},\omega ) \cdot \tens{G}(\B{r}'',\B{r}',\omega ),
\end{align}
where Lorentz reciprocity $\tens{G}(\B{r},\B{r}',\omega )=\tens{G}^\mathrm{T}(\B{r}',\B{r},\omega )$ has been used. Merit functions for observables that depend on $\nabla \B{G}(\B{r},\B{r}',\omega)$ can be obtained via the replacements $\tens{G}(\B{r},\B{r}',\omega)\mapsto\nabla \tens{G}(\B{r},\B{r}',\omega)$ and $\tens{G}^\mathrm{T}(\B{r}'',\B{r},\omega )\mapsto \tens{G}^\mathrm{T}(\B{r}'',\B{r},\omega )\overleftarrow{\nabla}$.

There are several features of \eqref{BasicDF} worth commenting on. In traditional presentations of adjoint optimisation, the equivalent of \eqref{BasicDF} is represented as the product of two electric fields. The first is the `direct' field, which is simply the electric field induced by the sources present in the system. The second is the adjoint field, which is that generated by a dipole oscillator at the observation point with an amplitude given by the electric-field derivative of the merit function. The advantage of adjoint methods is that the optimal value of the merit function can be found with only two simulations (rather than a brute force method entailing placement of a dielectric inclusion at each possible point in the optimisation region and repeatedly simulating for each). This is reflected our version of the merit function change shown in \eqref{BasicDF}; once the two independent Green's tensors for a source at $\B{r}'$ and a source at $\B{r}$ in a given environment (e.g. vacuum) have been calculated, $\delta F$ is known at all points. The link with the adjoint electric field is simply that one of the Green's tensors in \eqref{BasicDF} has been transposed. 

\begin{figure}
\includegraphics[width = \columnwidth]{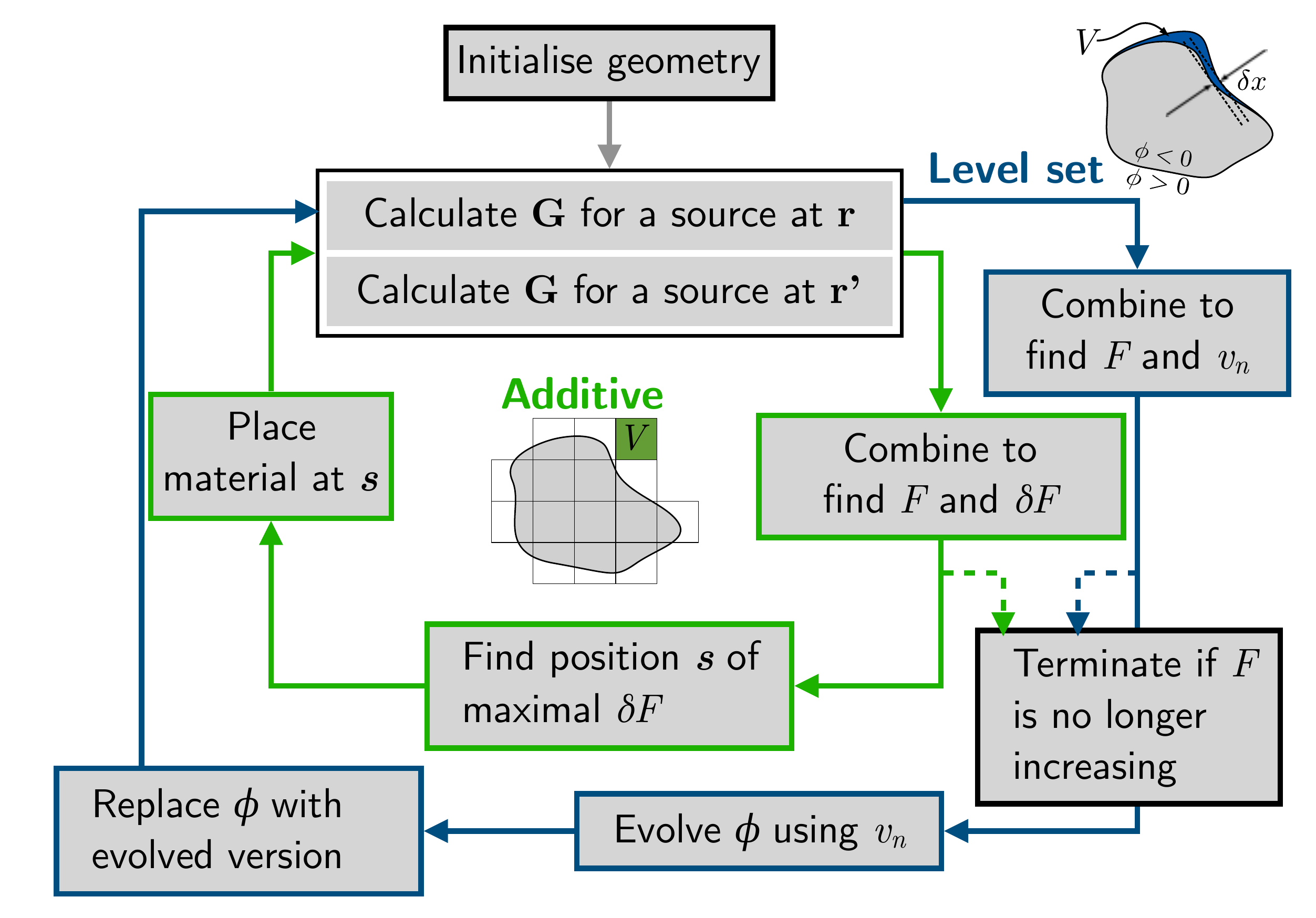}
\caption{Flow of the optimisation scheme, in either the level set (outer loop, blue) or additive approaches (inner loop, green). }
\label{ProgramFlow}
\end{figure}

At this point one has at least two choices for how to practically implement an optimisation --- the simplest is an additive scheme illustrated in \marker{the center of} Fig.~\ref{ProgramFlow}. Here a small block of material is added at the point of maximal $\delta F$, then the two Green's tensors in the new geometry are recalculated and combined to find a the next optimal point, an so on as indicated in Fig.~\ref{ProgramFlow}. 

The second way to implement the optimisation consists of gradually optimising the shape of an initial object by changing its boundary, known as the level-set method \cite{Osher1988}. This takes advantage of the fact that \eqref{BasicDF} avoids any explicit reference to electric or magnetic fields, thereby avoiding complications with the discontinuities usually found when the fields either side of boundary need to be considered, \marker{rather including them in the Green's tensor itself}. Here, the initial boundary shape (as well as its subsequent evolution) is encoded by a chosen function $\phi$. This is defined as negative inside the boundary, zero on it and positive outside, as indicated in Fig.~\ref{ProgramFlow}. 
Introducing a `time' parameter $t$ representing iteration, one is led to the following equation of motion governing the shape of the boundary \cite{Osher1988}:
$
\dot{\phi} (\B{r}(t),t) + v_n |\nabla \phi(\B{r}(t),t)| = 0
$
where $v_n$ is the velocity of motion normal to the surface. Formally, this is an advection equation which can be solved using techniques from fluid dynamics. Taking the volume $V$ in \eqref{BasicDF} to be that defined by the function $\phi$, we can let;
$
\int_V d^3 \B{r}'' \to \int_{\partial V} \d A \,  \delta x(\B{r}'')= \int_{\partial V} \d A \, v_n \delta t,
$
where the shape deformation has been assumed to be small, as can be ensured by a sufficiently small time step $\delta t$ in the evolution process. If the integrand of the $\B{r}''$ integral in \eqref{BasicDF} is positive at each iteration, the value of the merit function will continually increase. Positivity of \eqref{BasicDF} can then be ensured by using a velocity such that $ \partial F =  \int_{\partial V} \d A \, v_n^2 \delta t$, which means identifying;
\begin{align}
v_n &= 2 \text{Re} \int_0^\infty \!\!\!\! \d\omega \int \d^3\B{r} \! \int \d^3 \B{r}'  \alpha(\B{r}'') \notag \\
& \qquad \times \frac{\partial f}{\partial \tens{G}}(\B{r},\B{r}',\omega)\frob \tens{G}^\mathrm{T}(\B{r}'',\B{r},\omega ) \cdot \tens{G}(\B{r}'',\B{r}',\omega ),
\end{align}
This velocity can be directly calculated for a given $\B{G}$, then inserted into the advection equation, after which $\phi$ is evolved for $\delta t$. This delivers a new $\phi$, which defines a new geometry, for which we can calculate the new $\B{G}$ and the process iterates. 

Equation \eqref{BasicDF} can be directly applied to any quantity that can be expressed in terms of the Green's dyadic $\tens{G}$. This includes Casimir \cite{Casimir1948,Raabe2003} and Casimir-Polder forces \cite{Casimir1948a,Buhmann2004}, spontaneous decay (Purcell factor) \cite{Purcell1946,Joulain2003}, quantum friction \cite{Pendry1997,Klatt2017a}, interatomic Coulombic decay \cite{Cederbaum1997,Hemmerich2018}, radiative heat transfer \cite{Polder1971a,Volokitin2001}, van der Waals forces \cite{Buhmann2004b}, non-linear optical processes \cite{Lindel2019} and many more (the latter reference for each of these is where the formula in terms of $\tens{G}$ can be found). The merit functions for a selection of these are shown in Table \ref{MeritFnsTable}.

\begin{table*}[htbp]
\includegraphics[width =\textwidth]{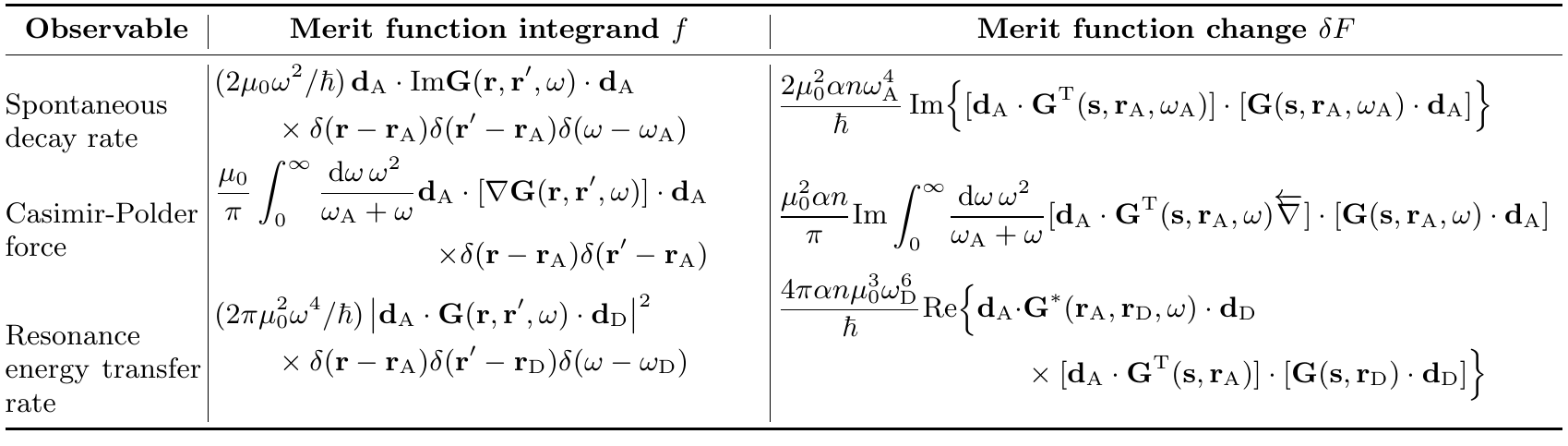}
  \caption{Non-exhaustive list of observables $f$ expressible in terms of $\B{G}$ and their associated merit function changes $\delta F$. In each case $\B{r}_\text{A}$ and $\B{r}_\text{D}$ are the positions of any atoms involved. The expression of the merit function change for spontaneous decay is equivalent to that used in Ref.~\cite{Liang2013}, as should be expected. }
  \label{MeritFnsTable}
\end{table*}

\paragraph*{Example implementation:} In order to demonstrate the application of \eqref{BasicDF}, we make some simplifying assumptions. We assume that the dielectric additions are homogenous and sufficiently small that the integral over $\B{r}''$ can be approximated by the value at its centre $\B{s}$:
\begin{align} \label{BasicDFS}
\delta F = 2 & \mu_0 \alpha n  \text{Re} \int \d\omega \int \d^3\B{r} \! \int \d^3 \B{r}' \notag  \\
&\times \frac{\partial F}{\partial \tens{G}}(\B{r},\B{r}',\omega)\frob\tens{G}^\text{T}(\B{s},\B{r},\omega ) \tens{G}(\B{s},\B{r}',\omega )\, . 
\end{align}
In practice, quantities which depend on the field at a single frequency \marker{and at a single position} are considerably more computationally tractable than their multi-frequency, \marker{bulk medium} counterparts. Here we concentrate on a simple and universal phenomenon which is well-approximated by \marker{radiation of} a single frequency \marker{interacting with a point-like atom} --- resonant energy transfer (RET). \marker{Extension to bulk media would not involve too much extra computational overhead since this would still only require two Green's tensors, but calculations of quantities dependent on a continuous spectrum (e.g. ground-state Casimir-Polder forces) would require re-calculation of $\tens{G}$ at very many frequencies. Our Green's tensor-based method can also be applied to nonlinear processes such as sum- or frequency-difference generation \cite{Lindel2019}} 

 We will work in the dipole approximation and aim to optimise the RET rate $\Gamma$ between dipole moments $\B{d}_\text{A}$ and $\B{d}_\text{D}$, meaning we take;
\begin{align}
f_\text{RET}[\B{G}(\B{r},\B{r}',\omega)] &=  \frac{2\pi \mu_0^2 \omega^4}{\hbar}\left| \B{d}_\text{A} \cdot \tens{G}(\B{r}, \B{r}',\omega)  \cdot \B{d}_\text{D} \right|^2 \notag \\
& \times \delta(\B{r}-\B{r}_\text{A}) \delta(\B{r}'-\B{r}_\text{D})  \delta(\omega - \omega_\text{D})
\end{align}
We then have simply;
\begin{equation}\label{FRET}
F_\text{RET}= \frac{2\pi \mu_0^2 \omega_\text{D}^4}{\hbar}  \left| \B{d}_\text{A} \cdot \tens{G}(\B{r}_\text{A}, \B{r}_\text{D},\omega)  \cdot \B{d}_\text{D} \right|^2 =\Gamma
\end{equation}
which is the well-known of resonance energy transfer rate $\Gamma$. Using this in \eqref{BasicDFS}, after some algebra one finds
\begin{align}\label{dFRet}
\delta F_\text{RET} &=  \frac{4\pi\alpha n \mu_0^3 \omega_\text{D}^4}{\hbar}  \text{Re} \Big\{   \B{d}_\text{A}   \tens{G}^*(\B{r}_\text{A}, \B{r}_\text{D},\omega)  \B{d}_\text{D} \notag \\
& \qquad   \times  [ \B{d}_\text{A} \tens{G}^\text{T}(\B{s},\B{r}_\text{A}) ]\cdot [ \tens{G}(\B{s},\B{r}_\text{D}) \B{d}_\text{D}] \Big\}
\end{align}
which is the equation we will work with from here on.

\begin{figure}
\centering
\includegraphics[width = 0.9\columnwidth]{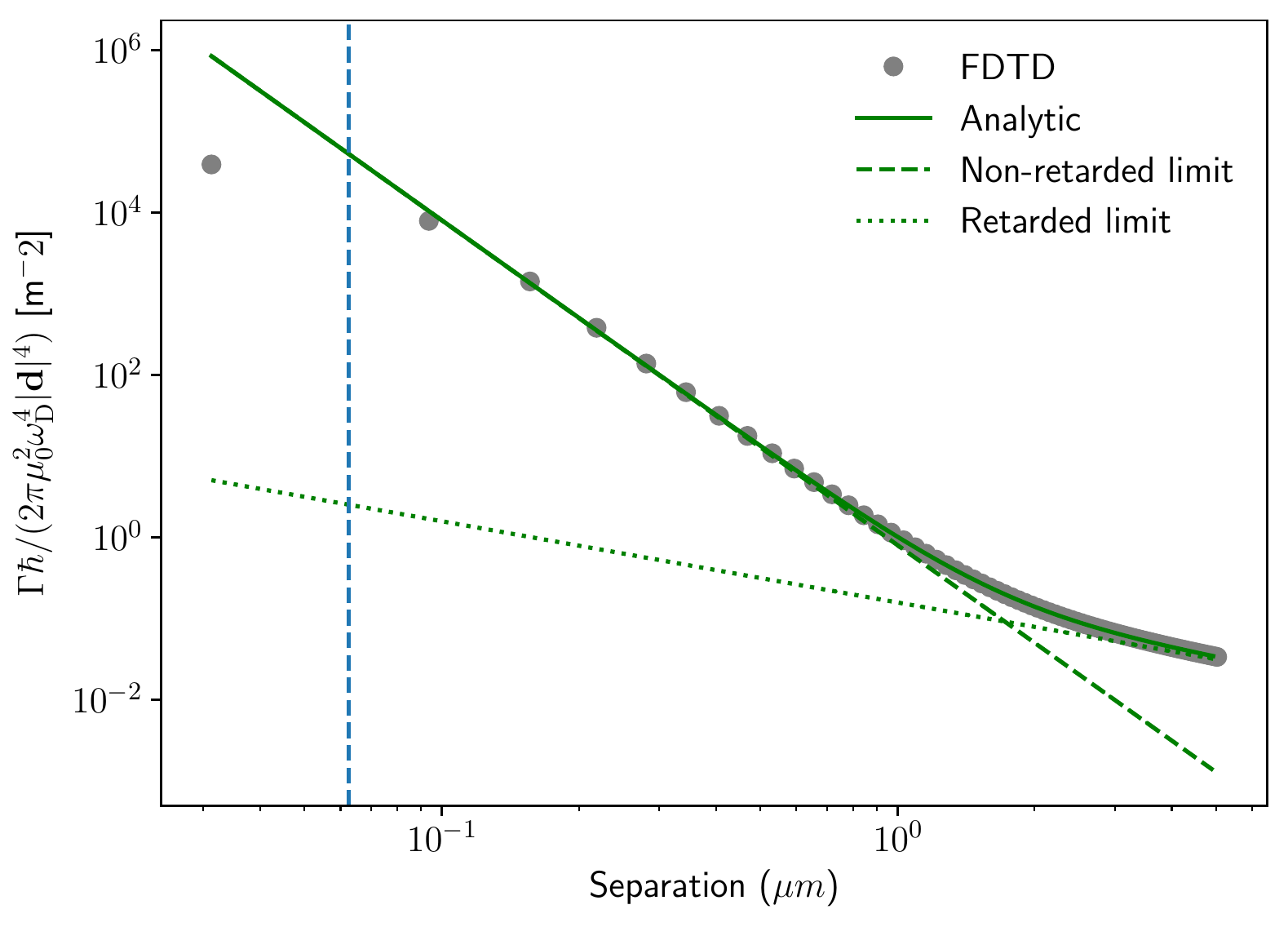}
\caption{Numerical vs analytic results for 2D-RET with donor and acceptor dipole moments parallel and of identical magnitude, with a transition wavelength of $500$nm.  Agreement is excellent until the interparticle distance approaches the pixel size used in this simulation, which was $(1/16)\mu \mathrm{m} = 62.5\mathrm{nm}$ as indicated by the vertical dashed line. All distances used in the rest of this work are well above this. }
\label{ValidationPlot}
\end{figure}

In order to demonstrate the main features of the method we restrict ourselves to systems with translational invariance along one axis, meaning they can be considered as effectively two-dimensional. In order to validate the two-dimensional RET results that we will calculate (as well as the general FDTD approach), it is necessary to have an analytic expression for RET in two dimensions. 
Formally, 2D-RET is equivalent to taking a pair of `line dipoles' each consisting of two infinitely extended parallel oppositely-charged wires in three dimensions, as discussed in detail in \cite{Martin1998}. The Green's tensor from  \cite{Martin1998} can be 
directly substituted into \eqref{FRET}, resulting in a lengthy expression,
which can be simplified by noting that in situations of practical interest the dipoles are often randomly oriented necessitating an isotropic average, which gives;
\begin{align}\label{Gamma2DIso}
\Gamma^\text{iso}_{\text{2D}} =\frac{2\pi \mu_0^2 \omega_\text{D}^4}{\hbar}  \frac{1}{16 \zeta } \bigg \{\Big[2 & \zeta  H_0^{(1)}(\zeta )- H_1^{(1)}(\zeta )\Big] H_0^{(2)}(\zeta )\notag \\
&+H_2^{(1)}(\zeta ) H_1^{(2)}(\zeta ) \bigg\}
\end{align}
where $H_n^{(1)}$ and $H_n^{(2)}$ are Hankel functions of the first and second kind respectively, and $\zeta = \omega_\text{D} \rho/c$. 
\eqref{Gamma2DIso} can be used to validate our general FDTD, see Fig.~\ref{ValidationPlot}. 

\begin{figure}[t!]
\includegraphics[width =\columnwidth]{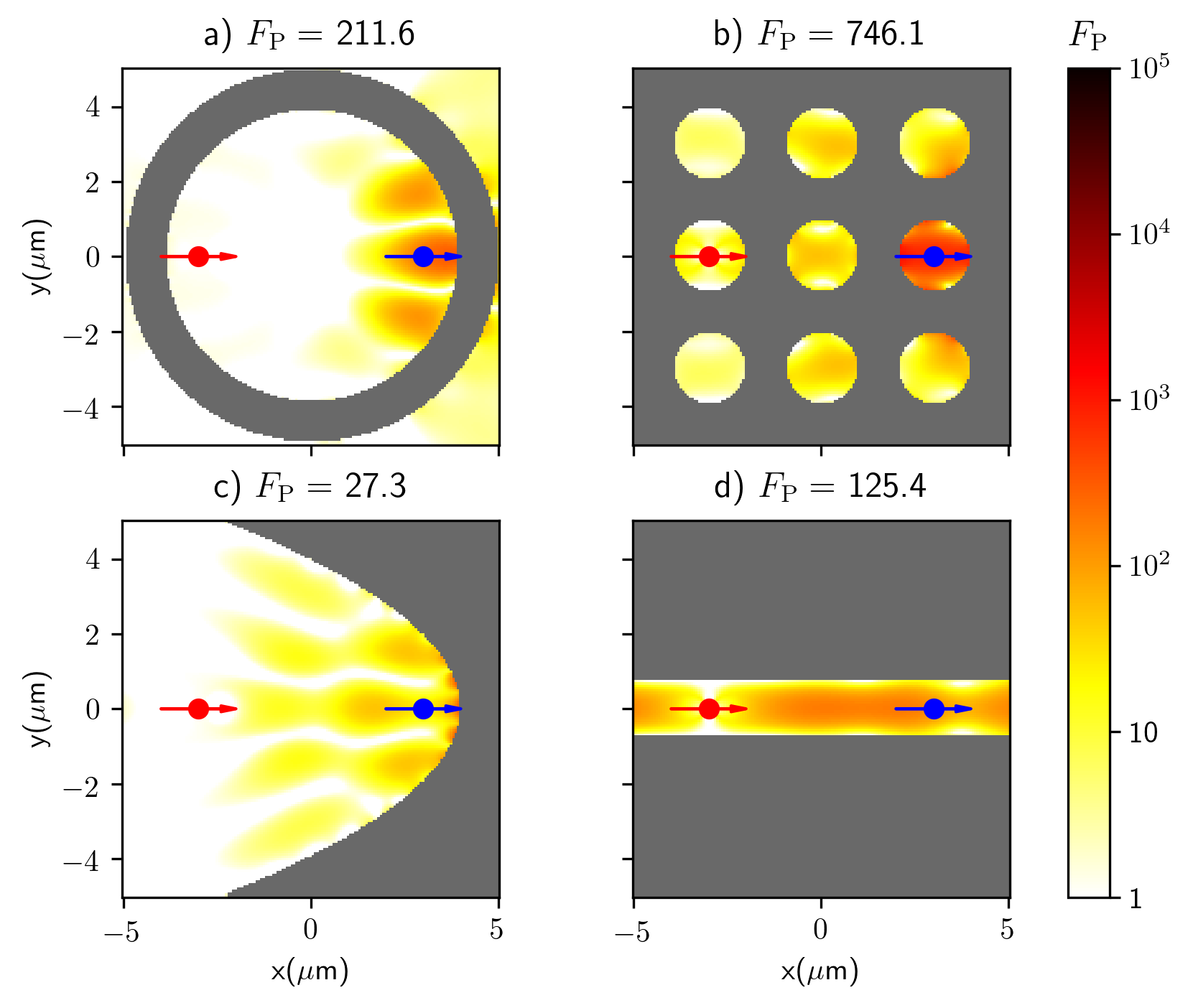}
\caption{RET in a) a ring-resonator, b) around a circle, c) with the donor at the focus of a parabola and d) in the center of a resonant (half-wavelength) cavity. \marker{The dipole on the left (red) is the donor which remains at a fixed position, while the other dipole (blue) is the acceptor which is allowed to move, with the arrows representing each dipole's spatial orientation. The position at which the acceptor is illustrated is that where the optimisation aims to enhance the RET rate. The grey regions have permittivity $\varepsilon = 12$, with the remainder being vacuum.} The simulations were done using a transition wavelength of $\pi\mu$m \marker{and a pixel size of 0.1$\mu$m.}}
\label{IntuitionPlots}
\end{figure}

We can now calculate the effect of arbitrary 2D geometries on RET by examining the dimensionless Purcell factor ${F_p ={\Gamma}/{\Gamma_0}}$, where $\Gamma_0$ is the rate in vacuum.  We initially choose some geometries which are expected to enhance RET, these are shown in Fig.~\ref{IntuitionPlots},
and give a maximum $F_p$ in the low hundreds. 

\begin{figure}[t!]
\includegraphics[width =\columnwidth]{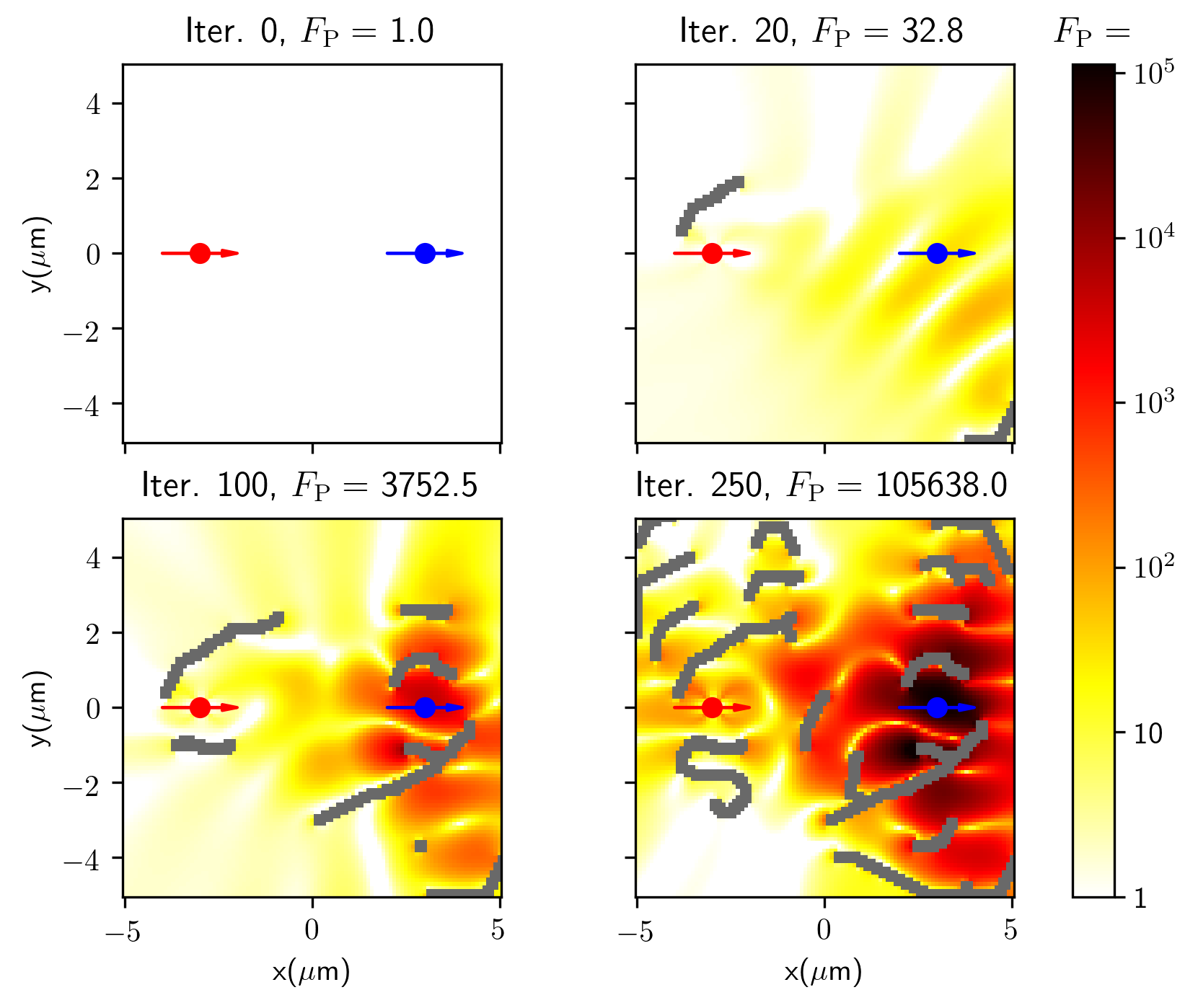}
\caption{Demonstration of dramatic enhancement of RET using the same parameters \marker{and colours} as in Fig.~\ref{IntuitionPlots}. The algorithm was constrained to not place any dielectric within 1$\mu$m of either the donor or acceptor.}
\label{IterativeEnhancement}
\end{figure}

Iterative optimisation techniques can improve on the examples chosen by hand. In order to demonstrate this, we use the additive approach shown in Fig.~\ref{ProgramFlow}, the results of which are shown in Fig.~\ref{IterativeEnhancement}.
An extremely large enhancement is found, reaching a factor of approximately $10^5$ after 250 iterations --- orders of magnitude higher than any enhancement found in the traditional designs shown in Fig.~\ref{IntuitionPlots}. It is worth noting that this extraordinarily high enhancement is achieved with a much smaller amount of dielectric material than in the traditional designs, even though the dielectric constant is identical. \marker{Finally we note that while all our results are scale-invariant (depending only on the ratio of the transition wavelength to the interatomic separation), the minimum feature size in our simulations is 0.1$\mu$m --- this is broadly consistent with the level to which complex structures can be manufactured (see, e.g. \cite{Su2018})}

In this Letter we have presented a convenient and system-agnostic version of adjoint optimisation of electromagnetism based entirely on the electromagnetic dyadic Green's tensor. This allows the techniques of inverse design to be applied to any of the vast number of interactions and processes which can be expressed in terms of this tensor. As an example we chose resonant energy transfer in two dimensions, showing orders of magnitude improvement in engineering potential compared to hand-made designs, while also deriving some new analytic results along the way. Extensions of our work could include application of the general level-set optimisation equation presented here, three-dimensional simulations and consideration of other observables including for example quantum yield of fluorescence processes. Inverse design of solar cells \cite{Alaeian2012} can now be conducted at a microscopic level by explicitly optimising each step in the energy transport chain, leading to large potential increases in efficiency.

The authors thank the Deutsche Forschungsgemeinschaft (grant BU 1803/3-1476), and R.B. acknowledges financial support by the Alexander von Humboldt Foundation. \\


\begin{thebibliography}{30}%
\makeatletter
\providecommand \@ifxundefined [1]{%
 \@ifx{#1\undefined}
}%
\providecommand \@ifnum [1]{%
 \ifnum #1\expandafter \@firstoftwo
 \else \expandafter \@secondoftwo
 \fi
}%
\providecommand \@ifx [1]{%
 \ifx #1\expandafter \@firstoftwo
 \else \expandafter \@secondoftwo
 \fi
}%
\providecommand \natexlab [1]{#1}%
\providecommand \enquote  [1]{``#1''}%
\providecommand \bibnamefont  [1]{#1}%
\providecommand \bibfnamefont [1]{#1}%
\providecommand \citenamefont [1]{#1}%
\providecommand \href@noop [0]{\@secondoftwo}%
\providecommand \href [0]{\begingroup \@sanitize@url \@href}%
\providecommand \@href[1]{\@@startlink{#1}\@@href}%
\providecommand \@@href[1]{\endgroup#1\@@endlink}%
\providecommand \@sanitize@url [0]{\catcode `\\12\catcode `\$12\catcode
  `\&12\catcode `\#12\catcode `\^12\catcode `\_12\catcode `\%12\relax}%
\providecommand \@@startlink[1]{}%
\providecommand \@@endlink[0]{}%
\providecommand \url  [0]{\begingroup\@sanitize@url \@url }%
\providecommand \@url [1]{\endgroup\@href {#1}{\urlprefix }}%
\providecommand \urlprefix  [0]{URL }%
\providecommand \Eprint [0]{\href }%
\providecommand \doibase [0]{http://dx.doi.org/}%
\providecommand \selectlanguage [0]{\@gobble}%
\providecommand \bibinfo  [0]{\@secondoftwo}%
\providecommand \bibfield  [0]{\@secondoftwo}%
\providecommand \translation [1]{[#1]}%
\providecommand \BibitemOpen [0]{}%
\providecommand \bibitemStop [0]{}%
\providecommand \bibitemNoStop [0]{.\EOS\space}%
\providecommand \EOS [0]{\spacefactor3000\relax}%
\providecommand \BibitemShut  [1]{\csname bibitem#1\endcsname}%
\let\auto@bib@innerbib\@empty
\bibitem [{\citenamefont {Spuhler}\ \emph {et~al.}(1998)\citenamefont
  {Spuhler}, \citenamefont {Offrein}, \citenamefont {Bona}, \citenamefont
  {Germann}, \citenamefont {Massarek},\ and\ \citenamefont
  {Erni}}]{Spuhler1998}%
  \BibitemOpen
  \bibfield  {author} {\bibinfo {author} {\bibfnamefont {M.}~\bibnamefont
  {Spuhler}}, \bibinfo {author} {\bibfnamefont {B.}~\bibnamefont {Offrein}},
  \bibinfo {author} {\bibfnamefont {G.-L.}\ \bibnamefont {Bona}}, \bibinfo
  {author} {\bibfnamefont {R.}~\bibnamefont {Germann}}, \bibinfo {author}
  {\bibfnamefont {I.}~\bibnamefont {Massarek}}, \ and\ \bibinfo {author}
  {\bibfnamefont {D.}~\bibnamefont {Erni}},\ }\href {\doibase
  10.1109/50.712252} {\bibfield  {journal} {\bibinfo  {journal} {J. Light.
  Technol.}\ }\textbf {\bibinfo {volume} {16}},\ \bibinfo {pages} {1680}
  (\bibinfo {year} {1998})}\BibitemShut {NoStop}%
\bibitem [{\citenamefont {Dobson}\ and\ \citenamefont
  {Cox}(1999)}]{Dobson1999}%
  \BibitemOpen
  \bibfield  {author} {\bibinfo {author} {\bibfnamefont {D.~C.}\ \bibnamefont
  {Dobson}}\ and\ \bibinfo {author} {\bibfnamefont {S.~J.}\ \bibnamefont
  {Cox}},\ }\href {\doibase 10.1137/S0036139998338455} {\bibfield  {journal}
  {\bibinfo  {journal} {SIAM J. Appl. Math.}\ }\textbf {\bibinfo {volume}
  {59}},\ \bibinfo {pages} {2108} (\bibinfo {year} {1999})}\BibitemShut
  {NoStop}%
\bibitem [{\citenamefont {Jameson}(1988)}]{Jameson1988}%
  \BibitemOpen
  \bibfield  {author} {\bibinfo {author} {\bibfnamefont {A.}~\bibnamefont
  {Jameson}},\ }\href {\doibase 10.1007/BF01061285} {\bibfield  {journal}
  {\bibinfo  {journal} {J. Sci. Comput.}\ }\textbf {\bibinfo {volume} {3}},\
  \bibinfo {pages} {233} (\bibinfo {year} {1988})}\BibitemShut {NoStop}%
\bibitem [{\citenamefont {Jensen}\ and\ \citenamefont
  {Sigmund}(2004)}]{Jensen2004}%
  \BibitemOpen
  \bibfield  {author} {\bibinfo {author} {\bibfnamefont {J.~S.}\ \bibnamefont
  {Jensen}}\ and\ \bibinfo {author} {\bibfnamefont {O.}~\bibnamefont
  {Sigmund}},\ }\href {\doibase 10.1063/1.1688450} {\bibfield  {journal}
  {\bibinfo  {journal} {Appl. Phys. Lett.}\ }\textbf {\bibinfo {volume} {84}},\
  \bibinfo {pages} {2022} (\bibinfo {year} {2004})}\BibitemShut {NoStop}%
\bibitem [{\citenamefont {Kao}\ \emph {et~al.}(2005)\citenamefont {Kao},
  \citenamefont {Osher},\ and\ \citenamefont {Yablonovitch}}]{Kao2005}%
  \BibitemOpen
  \bibfield  {author} {\bibinfo {author} {\bibfnamefont {C.~Y.}\ \bibnamefont
  {Kao}}, \bibinfo {author} {\bibfnamefont {S.}~\bibnamefont {Osher}}, \ and\
  \bibinfo {author} {\bibfnamefont {E.}~\bibnamefont {Yablonovitch}},\ }\href
  {\doibase 10.1007/s00340-005-1877-3} {\bibfield  {journal} {\bibinfo
  {journal} {Appl. Phys. B}\ }\textbf {\bibinfo {volume} {81}},\ \bibinfo
  {pages} {235} (\bibinfo {year} {2005})}\BibitemShut {NoStop}%
\bibitem [{\citenamefont {Alaeian}\ \emph {et~al.}(2012)\citenamefont
  {Alaeian}, \citenamefont {Atre},\ and\ \citenamefont {Dionne}}]{Alaeian2012}%
  \BibitemOpen
  \bibfield  {author} {\bibinfo {author} {\bibfnamefont {H.}~\bibnamefont
  {Alaeian}}, \bibinfo {author} {\bibfnamefont {A.~C.}\ \bibnamefont {Atre}}, \
  and\ \bibinfo {author} {\bibfnamefont {J.~A.}\ \bibnamefont {Dionne}},\
  }\href {\doibase 10.1088/2040-8978/14/2/024006} {\bibfield  {journal}
  {\bibinfo  {journal} {J. Opt.}\ }\textbf {\bibinfo {volume} {14}},\ \bibinfo
  {pages} {024006} (\bibinfo {year} {2012})}\BibitemShut {NoStop}%
\bibitem [{\citenamefont {Piggott}\ \emph {et~al.}(2015)\citenamefont
  {Piggott}, \citenamefont {Lu}, \citenamefont {Lagoudakis}, \citenamefont
  {Petykiewicz}, \citenamefont {Babinec},\ and\ \citenamefont
  {Vuckovi{\'{c}}}}]{Piggott2015}%
  \BibitemOpen
  \bibfield  {author} {\bibinfo {author} {\bibfnamefont {A.~Y.}\ \bibnamefont
  {Piggott}}, \bibinfo {author} {\bibfnamefont {J.}~\bibnamefont {Lu}},
  \bibinfo {author} {\bibfnamefont {K.~G.}\ \bibnamefont {Lagoudakis}},
  \bibinfo {author} {\bibfnamefont {J.}~\bibnamefont {Petykiewicz}}, \bibinfo
  {author} {\bibfnamefont {T.~M.}\ \bibnamefont {Babinec}}, \ and\ \bibinfo
  {author} {\bibfnamefont {J.}~\bibnamefont {Vuckovi{\'{c}}}},\ }\href
  {\doibase 10.1038/nphoton.2015.69} {\bibfield  {journal} {\bibinfo  {journal}
  {Nat. Photonics}\ }\textbf {\bibinfo {volume} {9}},\ \bibinfo {pages} {374}
  (\bibinfo {year} {2015})}\BibitemShut {NoStop}%
\bibitem [{\citenamefont {Jensen}\ and\ \citenamefont
  {Sigmund}(2011)}]{Jensen2011}%
  \BibitemOpen
  \bibfield  {author} {\bibinfo {author} {\bibfnamefont {J.}~\bibnamefont
  {Jensen}}\ and\ \bibinfo {author} {\bibfnamefont {O.}~\bibnamefont
  {Sigmund}},\ }\href {\doibase 10.1002/lpor.201000014} {\bibfield  {journal}
  {\bibinfo  {journal} {Laser Photon. Rev.}\ }\textbf {\bibinfo {volume} {5}},\
  \bibinfo {pages} {308} (\bibinfo {year} {2011})}\BibitemShut {NoStop}%
\bibitem [{\citenamefont {Molesky}\ \emph {et~al.}(2018)\citenamefont
  {Molesky}, \citenamefont {Lin}, \citenamefont {Piggott}, \citenamefont {Jin},
  \citenamefont {Vuckovi{\'{c}}},\ and\ \citenamefont
  {Rodriguez}}]{Molesky2018}%
  \BibitemOpen
  \bibfield  {author} {\bibinfo {author} {\bibfnamefont {S.}~\bibnamefont
  {Molesky}}, \bibinfo {author} {\bibfnamefont {Z.}~\bibnamefont {Lin}},
  \bibinfo {author} {\bibfnamefont {A.~Y.}\ \bibnamefont {Piggott}}, \bibinfo
  {author} {\bibfnamefont {W.}~\bibnamefont {Jin}}, \bibinfo {author}
  {\bibfnamefont {J.}~\bibnamefont {Vuckovi{\'{c}}}}, \ and\ \bibinfo {author}
  {\bibfnamefont {A.~W.}\ \bibnamefont {Rodriguez}},\ }\href {\doibase
  10.1038/s41566-018-0246-9} {\bibfield  {journal} {\bibinfo  {journal} {Nat.
  Photonics}\ }\textbf {\bibinfo {volume} {12}},\ \bibinfo {pages} {659}
  (\bibinfo {year} {2018})}\BibitemShut {NoStop}%
\bibitem [{\citenamefont {Casimir}\ and\ \citenamefont
  {Polder}(1948)}]{Casimir1948a}%
  \BibitemOpen
  \bibfield  {author} {\bibinfo {author} {\bibfnamefont {H.~B.~G.}\
  \bibnamefont {Casimir}}\ and\ \bibinfo {author} {\bibfnamefont
  {D.}~\bibnamefont {Polder}},\ }\href {\doibase 10.1103/PhysRev.73.360}
  {\bibfield  {journal} {\bibinfo  {journal} {Phys. Rev.}\ }\textbf {\bibinfo
  {volume} {73}},\ \bibinfo {pages} {360} (\bibinfo {year} {1948})}\BibitemShut
  {NoStop}%
\bibitem [{\citenamefont {F{\"{o}}rster}(1948)}]{Forster1948}%
  \BibitemOpen
  \bibfield  {author} {\bibinfo {author} {\bibfnamefont {T.}~\bibnamefont
  {F{\"{o}}rster}},\ }\href {\doibase 10.1002/andp.19484370105} {\bibfield
  {journal} {\bibinfo  {journal} {Ann. Phys.}\ }\textbf {\bibinfo {volume}
  {437}},\ \bibinfo {pages} {55} (\bibinfo {year} {1948})}\BibitemShut
  {NoStop}%
\bibitem [{\citenamefont {Gruner}\ and\ \citenamefont
  {Welsch}(1996)}]{Gruner1996a}%
  \BibitemOpen
  \bibfield  {author} {\bibinfo {author} {\bibfnamefont {T.}~\bibnamefont
  {Gruner}}\ and\ \bibinfo {author} {\bibfnamefont {D.-G.}\ \bibnamefont
  {Welsch}},\ }\href {\doibase 10.1103/PhysRevA.53.1818} {\bibfield  {journal}
  {\bibinfo  {journal} {Phys. Rev. A}\ }\textbf {\bibinfo {volume} {53}},\
  \bibinfo {pages} {1818} (\bibinfo {year} {1996})}\BibitemShut {NoStop}%
\bibitem [{Note1()}]{Note1}%
  \BibitemOpen
  \bibinfo {note} {These are usually called $\protect \mathbf {f}$ and
  $\protect \mathbf {f}^\dagger $ in macroscopic QED, here we avoid that
  notation in order to avoid confusion with the merit functions $f$ introduced
  later.}\BibitemShut {Stop}%
\bibitem [{\citenamefont {Osher}\ and\ \citenamefont
  {Sethian}(1988)}]{Osher1988}%
  \BibitemOpen
  \bibfield  {author} {\bibinfo {author} {\bibfnamefont {S.}~\bibnamefont
  {Osher}}\ and\ \bibinfo {author} {\bibfnamefont {J.~A.}\ \bibnamefont
  {Sethian}},\ }\href {\doibase 10.1016/0021-9991(88)90002-2} {\bibfield
  {journal} {\bibinfo  {journal} {J. Comput. Phys.}\ }\textbf {\bibinfo
  {volume} {79}},\ \bibinfo {pages} {12} (\bibinfo {year} {1988})}\BibitemShut
  {NoStop}%
\bibitem [{\citenamefont {Casimir}(1948)}]{Casimir1948}%
  \BibitemOpen
  \bibfield  {author} {\bibinfo {author} {\bibfnamefont {H.~B.~G.}\
  \bibnamefont {Casimir}},\ }\href {\doibase citeulike-article-id:8810715}
  {\bibfield  {journal} {\bibinfo  {journal} {Proc. K. Ned. Akad.}\ }\textbf
  {\bibinfo {volume} {360}},\ \bibinfo {pages} {793} (\bibinfo {year}
  {1948})}\BibitemShut {NoStop}%
\bibitem [{\citenamefont {Raabe}\ \emph {et~al.}(2003)\citenamefont {Raabe},
  \citenamefont {Kn{\"{o}}ll},\ and\ \citenamefont {Welsch}}]{Raabe2003}%
  \BibitemOpen
  \bibfield  {author} {\bibinfo {author} {\bibfnamefont {C.}~\bibnamefont
  {Raabe}}, \bibinfo {author} {\bibfnamefont {L.}~\bibnamefont {Kn{\"{o}}ll}},
  \ and\ \bibinfo {author} {\bibfnamefont {D.-G.}\ \bibnamefont {Welsch}},\
  }\href {\doibase 10.1103/PhysRevA.68.033810} {\bibfield  {journal} {\bibinfo
  {journal} {Phys. Rev. A}\ }\textbf {\bibinfo {volume} {68}},\ \bibinfo
  {pages} {033810} (\bibinfo {year} {2003})}\BibitemShut {NoStop}%
\bibitem [{\citenamefont {Buhmann}\ \emph
  {et~al.}(2004{\natexlab{a}})\citenamefont {Buhmann}, \citenamefont
  {Kn{\"{o}}ll}, \citenamefont {Welsch},\ and\ \citenamefont
  {Dung}}]{Buhmann2004}%
  \BibitemOpen
  \bibfield  {author} {\bibinfo {author} {\bibfnamefont {S.~Y.}\ \bibnamefont
  {Buhmann}}, \bibinfo {author} {\bibfnamefont {L.}~\bibnamefont
  {Kn{\"{o}}ll}}, \bibinfo {author} {\bibfnamefont {D.-G.}\ \bibnamefont
  {Welsch}}, \ and\ \bibinfo {author} {\bibfnamefont {H.~T.}\ \bibnamefont
  {Dung}},\ }\href {\doibase 10.1103/PhysRevA.70.052117} {\bibfield  {journal}
  {\bibinfo  {journal} {Phys. Rev. A}\ }\textbf {\bibinfo {volume} {70}},\
  \bibinfo {pages} {052117} (\bibinfo {year} {2004}{\natexlab{a}})}\BibitemShut
  {NoStop}%
\bibitem [{\citenamefont {Purcell}(1946)}]{Purcell1946}%
  \BibitemOpen
  \bibfield  {author} {\bibinfo {author} {\bibfnamefont {E.~M.}\ \bibnamefont
  {Purcell}},\ }\href {\doibase 10.1103/PhysRev.69.674.2} {\bibfield  {journal}
  {\bibinfo  {journal} {Proc. Am. Phys. Soc.}\ }\textbf {\bibinfo {volume}
  {69}},\ \bibinfo {pages} {674} (\bibinfo {year} {1946})}\BibitemShut
  {NoStop}%
\bibitem [{\citenamefont {Joulain}\ \emph {et~al.}(2003)\citenamefont
  {Joulain}, \citenamefont {Carminati}, \citenamefont {Mulet},\ and\
  \citenamefont {Greffet}}]{Joulain2003}%
  \BibitemOpen
  \bibfield  {author} {\bibinfo {author} {\bibfnamefont {K.}~\bibnamefont
  {Joulain}}, \bibinfo {author} {\bibfnamefont {R.}~\bibnamefont {Carminati}},
  \bibinfo {author} {\bibfnamefont {J.-P.}\ \bibnamefont {Mulet}}, \ and\
  \bibinfo {author} {\bibfnamefont {J.-J.}\ \bibnamefont {Greffet}},\ }\href
  {\doibase 10.1103/PhysRevB.68.245405} {\bibfield  {journal} {\bibinfo
  {journal} {Phys. Rev. B}\ }\textbf {\bibinfo {volume} {68}},\ \bibinfo
  {pages} {245405} (\bibinfo {year} {2003})}\BibitemShut {NoStop}%
\bibitem [{\citenamefont {Pendry}(1997)}]{Pendry1997}%
  \BibitemOpen
  \bibfield  {author} {\bibinfo {author} {\bibfnamefont {J.~B.}\ \bibnamefont
  {Pendry}},\ }\href {\doibase 10.1088/0953-8984/9/47/001} {\bibfield
  {journal} {\bibinfo  {journal} {J. Phys. Condens. Matter}\ }\textbf {\bibinfo
  {volume} {9}},\ \bibinfo {pages} {10301} (\bibinfo {year}
  {1997})}\BibitemShut {NoStop}%
\bibitem [{\citenamefont {Klatt}\ \emph {et~al.}(2017)\citenamefont {Klatt},
  \citenamefont {Farias}, \citenamefont {Dalvit},\ and\ \citenamefont
  {Buhmann}}]{Klatt2017a}%
  \BibitemOpen
  \bibfield  {author} {\bibinfo {author} {\bibfnamefont {J.}~\bibnamefont
  {Klatt}}, \bibinfo {author} {\bibfnamefont {M.~B.}\ \bibnamefont {Farias}},
  \bibinfo {author} {\bibfnamefont {D.~A.~R.}\ \bibnamefont {Dalvit}}, \ and\
  \bibinfo {author} {\bibfnamefont {S.~Y.}\ \bibnamefont {Buhmann}},\ }\href
  {http://link.aps.org/doi/10.1103/PhysRevA.95.052510} {\bibfield  {journal}
  {\bibinfo  {journal} {Phys. Rev. A}\ }\textbf {\bibinfo {volume} {95}},\
  \bibinfo {pages} {052510} (\bibinfo {year} {2017})}\BibitemShut {NoStop}%
\bibitem [{\citenamefont {Cederbaum}\ \emph {et~al.}(1997)\citenamefont
  {Cederbaum}, \citenamefont {Zobeley},\ and\ \citenamefont
  {Tarantelli}}]{Cederbaum1997}%
  \BibitemOpen
  \bibfield  {author} {\bibinfo {author} {\bibfnamefont {L.~S.}\ \bibnamefont
  {Cederbaum}}, \bibinfo {author} {\bibfnamefont {J.}~\bibnamefont {Zobeley}},
  \ and\ \bibinfo {author} {\bibfnamefont {F.}~\bibnamefont {Tarantelli}},\
  }\href {\doibase 10.1103/PhysRevLett.79.4778} {\bibfield  {journal} {\bibinfo
   {journal} {Phys. Rev. Lett.}\ }\textbf {\bibinfo {volume} {79}},\ \bibinfo
  {pages} {4778} (\bibinfo {year} {1997})}\BibitemShut {NoStop}%
\bibitem [{\citenamefont {Hemmerich}\ \emph {et~al.}(2018)\citenamefont
  {Hemmerich}, \citenamefont {Bennett},\ and\ \citenamefont
  {Buhmann}}]{Hemmerich2018}%
  \BibitemOpen
  \bibfield  {author} {\bibinfo {author} {\bibfnamefont {J.~L.}\ \bibnamefont
  {Hemmerich}}, \bibinfo {author} {\bibfnamefont {R.}~\bibnamefont {Bennett}},
  \ and\ \bibinfo {author} {\bibfnamefont {S.~Y.}\ \bibnamefont {Buhmann}},\
  }\href {\doibase 10.1038/s41467-018-05091-x} {\bibfield  {journal} {\bibinfo
  {journal} {Nat. Commun.}\ }\textbf {\bibinfo {volume} {9}},\ \bibinfo {pages}
  {2934} (\bibinfo {year} {2018})}\BibitemShut {NoStop}%
\bibitem [{\citenamefont {Polder}\ and\ \citenamefont {{Van
  Hove}}(1971)}]{Polder1971a}%
  \BibitemOpen
  \bibfield  {author} {\bibinfo {author} {\bibfnamefont {D.}~\bibnamefont
  {Polder}}\ and\ \bibinfo {author} {\bibfnamefont {M.}~\bibnamefont {{Van
  Hove}}},\ }\href {\doibase 10.1103/PhysRevB.4.3303} {\bibfield  {journal}
  {\bibinfo  {journal} {Phys. Rev. B}\ }\textbf {\bibinfo {volume} {4}},\
  \bibinfo {pages} {3303} (\bibinfo {year} {1971})},\ \Eprint
  {http://arxiv.org/abs/PhysRevB.4.3303} {arXiv:PhysRevB.4.3303 [10.1103]}
  \BibitemShut {NoStop}%
\bibitem [{\citenamefont {Volokitin}\ and\ \citenamefont
  {Persson}(2001)}]{Volokitin2001}%
  \BibitemOpen
  \bibfield  {author} {\bibinfo {author} {\bibfnamefont {A.~I.}\ \bibnamefont
  {Volokitin}}\ and\ \bibinfo {author} {\bibfnamefont {B.~N.~J.}\ \bibnamefont
  {Persson}},\ }\href {\doibase 10.1103/PhysRevB.63.205404} {\bibfield
  {journal} {\bibinfo  {journal} {Phys. Rev. B}\ }\textbf {\bibinfo {volume}
  {63}},\ \bibinfo {pages} {205404} (\bibinfo {year} {2001})}\BibitemShut
  {NoStop}%
\bibitem [{\citenamefont {Buhmann}\ \emph
  {et~al.}(2004{\natexlab{b}})\citenamefont {Buhmann}, \citenamefont {Dung},\
  and\ \citenamefont {Welsch}}]{Buhmann2004b}%
  \BibitemOpen
  \bibfield  {author} {\bibinfo {author} {\bibfnamefont {S.~Y.}\ \bibnamefont
  {Buhmann}}, \bibinfo {author} {\bibfnamefont {H.~T.}\ \bibnamefont {Dung}}, \
  and\ \bibinfo {author} {\bibfnamefont {D.-G.}\ \bibnamefont {Welsch}},\
  }\href {\doibase 10.1088/1464-4266/6/3/020} {\bibfield  {journal} {\bibinfo
  {journal} {J. Opt. B Quantum Semiclassical Opt.}\ }\textbf {\bibinfo {volume}
  {6}},\ \bibinfo {pages} {S127} (\bibinfo {year}
  {2004}{\natexlab{b}})}\BibitemShut {NoStop}%
\bibitem [{\citenamefont {Lindel}\ \emph {et~al.}(2019)\citenamefont {Lindel},
  \citenamefont {Bennett},\ and\ \citenamefont {Buhmann}}]{Lindel2019}%
  \BibitemOpen
  \bibfield  {author} {\bibinfo {author} {\bibfnamefont {F.}~\bibnamefont
  {Lindel}}, \bibinfo {author} {\bibfnamefont {R.}~\bibnamefont {Bennett}}, \
  and\ \bibinfo {author} {\bibfnamefont {S.~Y.}\ \bibnamefont {Buhmann}},\
  }\href {http://arxiv.org/abs/1905.10200} {\bibfield  {journal} {\bibinfo
  {journal} {arXiv quant-ph: 1905.10200}\ } (\bibinfo {year}
  {2019})}\BibitemShut {NoStop}%
\bibitem [{\citenamefont {Liang}\ and\ \citenamefont
  {Johnson}(2013)}]{Liang2013}%
  \BibitemOpen
  \bibfield  {author} {\bibinfo {author} {\bibfnamefont {X.}~\bibnamefont
  {Liang}}\ and\ \bibinfo {author} {\bibfnamefont {S.~G.}\ \bibnamefont
  {Johnson}},\ }\href {\doibase 10.1364/oe.21.030812} {\bibfield  {journal}
  {\bibinfo  {journal} {Opt. Express}\ }\textbf {\bibinfo {volume} {21}},\
  \bibinfo {pages} {30812} (\bibinfo {year} {2013})}\BibitemShut {NoStop}%
\bibitem [{\citenamefont {Martin}\ and\ \citenamefont
  {Piller}(1998)}]{Martin1998}%
  \BibitemOpen
  \bibfield  {author} {\bibinfo {author} {\bibfnamefont {O.~J.~F.}\
  \bibnamefont {Martin}}\ and\ \bibinfo {author} {\bibfnamefont {N.~B.}\
  \bibnamefont {Piller}},\ }\href {\doibase 10.1103/PhysRevE.58.3909}
  {\bibfield  {journal} {\bibinfo  {journal} {Phys. Rev. E}\ }\textbf {\bibinfo
  {volume} {58}},\ \bibinfo {pages} {3909} (\bibinfo {year}
  {1998})}\BibitemShut {NoStop}%
\bibitem [{\citenamefont {Su}\ \emph {et~al.}(2018)\citenamefont {Su},
  \citenamefont {Piggott}, \citenamefont {Sapra}, \citenamefont {Petykiewicz},\
  and\ \citenamefont {Vu{\v{c}}kovi{\'{c}}}}]{Su2018}%
  \BibitemOpen
  \bibfield  {author} {\bibinfo {author} {\bibfnamefont {L.}~\bibnamefont
  {Su}}, \bibinfo {author} {\bibfnamefont {A.~Y.}\ \bibnamefont {Piggott}},
  \bibinfo {author} {\bibfnamefont {N.~V.}\ \bibnamefont {Sapra}}, \bibinfo
  {author} {\bibfnamefont {J.}~\bibnamefont {Petykiewicz}}, \ and\ \bibinfo
  {author} {\bibfnamefont {J.}~\bibnamefont {Vu{\v{c}}kovi{\'{c}}}},\ }\href
  {\doibase 10.1021/acsphotonics.7b00987} {\bibfield  {journal} {\bibinfo
  {journal} {ACS Photonics}\ }\textbf {\bibinfo {volume} {5}},\ \bibinfo
  {pages} {301} (\bibinfo {year} {2018})}\BibitemShut {NoStop}%
\end{thebibliography}
\end{document}